\documentclass[a4paper]{article}

\usepackage{INTERSPEECH2020}

\usepackage{svg}
\usepackage{amsmath}
\usepackage{graphicx}
\usepackage{xcolor}
\usepackage{enumerate}
\usepackage{diagbox}
\usepackage{enumitem}
\usepackage{multirow}
\usepackage{comment}
\usepackage{lipsum}
\usepackage{booktabs}
\usepackage[super]{nth}

\title{DARTS-ASR: Differentiable Architecture Search for \\Multilingual Speech Recognition and Adaptation}
\name{Yi-Chen Chen$^{12}$, Jui-Yang Hsu$^1$, Cheng-Kuang Lee$^2$, Hung-yi Lee$^1$}
\address{
  $^1$Graduate Institute of Communication Engineering, National Taiwan University\\
  $^2$NVIDIA AI Technology Center, NVIDIA}
\email{\{f06942069,r07921053,hungyilee\}@ntu.edu.tw, ckl@nvidia.com}

\begin{document}

\maketitle
\begin{abstract}
In previous works, only parameter weights of ASR models are optimized under fixed-topology architecture.
However, the design of successful model architecture has always relied on human experience and intuition.
Besides, many hyperparameters related to model architecture need to be manually tuned.
Therefore in this paper, we propose an ASR approach with efficient gradient-based architecture search, DARTS-ASR.
In order to examine the generalizability of DARTS-ASR, we apply our approach not only on many languages to perform monolingual ASR, but also on a multilingual ASR setting.
Following previous works, we conducted experiments on a multilingual dataset, IARPA BABEL.
The experiment results show that our approach outperformed the baseline fixed-topology architecture by 10.2\% and 10.0\% relative reduction on character error rates under monolingual and multilingual ASR settings respectively.
Furthermore, we perform some analysis on the searched architectures by DARTS-ASR.
\end{abstract}

\section{Introduction}
\label{sec:intro}

Recently deep neural network (DNN) models have achieved huge success in many applications. 
A lot of empirical evidence has shown that network architecture matters significantly in fields like image classification (from AlexNet~\cite{krizhevsky2012imagenet} to ResNet~\cite{he2016deep}) or natural language processing (NLP) (Transformer~\cite{devlin2019bert}).
Despite the success of these DNNs, the architecture is still hard to design.
The popular architectures were usually invented and tuned by experts through a long process of trial and error.

For example, convolutional neural networks (CNN)~\cite{krizhevsky2012imagenet} have been proved to be more effective in image recognition tasks than DNNs with fully-connected layers.
CNNs were inspired by biological processes where the connectivity pattern between neurons resembles the organization of the animal visual cortex~\cite{fukushima1980neocognitron}. 
However, the birth of such successful model architecture always relies on human wisdom and a flash of insight.
Besides, many hyperparameters in CNNs still have to be carefully tuned, such as channel numbers, kernel sizes, strides, padding, pooling and activation functions for each layer.
Therefore, it is highly appealing to have an effective algorithm to discover and design architectures of DNNs automatically.

Many researchers have focused on automatic neural architecture search (NAS) algorithms, aiming to optimize not only parameter weights of a fixed-topology neural network architecture, but also the design of architecture itself.
Some approaches~\cite{DBLP:conf/iclr/ZophL17,zoph2018learning} use reinforcement learning (RL) to search for building blocks used in CNN. Some other approaches~\cite{real2019regularized} use evolutionary algorithms to find building blocks through mutation and tournament selection.
Some recent works also incorporate NAS into their approaches to speech recognition~\cite{baruwa2019leveraging} or keyword spotting~\cite{veniat2019stochastic,mazzawi2019improving}.
Although these approaches have achieved convincing results on many benchmark datasets, a huge amount of computational resources are needed to perform exploration in a search space.

Differentiable ARchiTecture Search (DARTS)~\cite{liu2018darts} uses a gradient-based method for efficient architecture search.
Instead of searching over discrete architecture candidates, with a continuous relaxation of architecture representation, the architecture can be jointly optimized with parameter weights directly by gradient descent.
On many benchmark datasets of image classification, more recent approaches~\cite{wu2019fbnet,xie2018snas,chen2019progressive} based on DARTS have discovered model architectures that achieved state-of-the-art results with similar parameter size to other models.

\begin{figure}[t]
  \centering
  \centerline{\includegraphics[width=\linewidth]{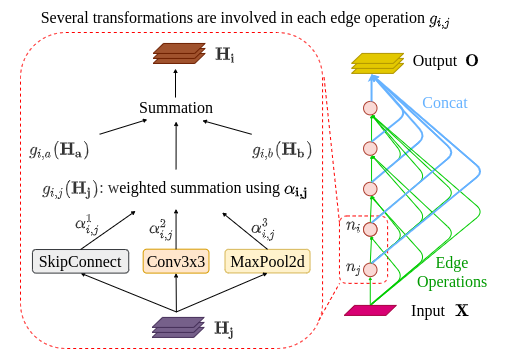}}
\caption{Differentiable ARchiTecture Search (DARTS) for ASR.}
\label{fig:DARTS-ASR}
\vspace{-0.5cm}
\end{figure}

Inspired by DARTS, in this paper we propose an ASR approach with efficient gradient-based architecture search, \textbf{DARTS-ASR}.
In order to examine the generalizability of DARTS-ASR, we apply our approach not only on many languages to perform monolingual ASR, but also on a multilingual ASR setting, where the architecture and parameter weights are pre-trained on some source languages, and then adapted on the target language.
It has recently been shown that multilingual ASR~\cite{vu2014multilingual,tong2017investigation,cho2018multilingual,tong2017multilingual,yi2018adversarial,adams2019massively,sanabria2018hierarchical,dalmia2018sequence,hsu2019meta} can improve ASR performance on many low-resource languages.
In the above previous works, the initial parameters or shared encoder learned from many source languages are used to build a better acoustic model for the target language.
Different from previous works, DARTS-ASR further learns better network architecture from the source languages. 

Following the previous works~\cite{cho2018multilingual,yi2018adversarial,dalmia2018sequence,hsu2019meta}, we conducted experiments on the multilingual dataset, IARPA BABEL~\cite{gales2014speech}.
The experiment results show that our approach outperformed the baseline fixed-topology architecture by 10.2\% and 10.0\% relative reduction on character error rates (CER) under monolingual and multilingual ASR settings respectively.
Furthermore, we perform some analysis on the searched architectures by DARTS-ASR.

\section{Proposed Approach: DARTS-ASR}
\label{sec:approach}

\begin{figure}[t]
\begin{minipage}[t]{0.54\linewidth}
  \centering
  \centerline{\includegraphics[width=\linewidth]{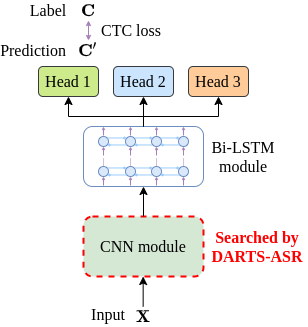}}
  \centerline{(a) The framework of ASR model.}\medskip
  \label{minipage:ASR} 
\end{minipage}
\hfill
\begin{minipage}[t]{0.45\linewidth}
  \centering
  \centerline{\includegraphics[width=0.5\linewidth]{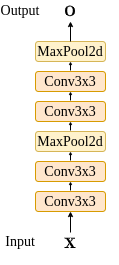}}
  \centerline{(b) CNN module as VGG.}\medskip
  \label{minipage:VGG} 
\end{minipage}
\vspace{-0.4cm}
\caption{Multilingual ASR model with CTC.}
\label{fig:model}
\vspace{-0.4cm}
\end{figure}

In previous works of ASR, network architectures were manually designed with human experience, and parameter weights can only be optimized under the fixed topology.
Although those networks work well in previous works, they are very likely not the optimal architectures for ASR.
In this paper, we propose \textbf{DARTS-ASR}, where the network architecture can be automatically learned jointly with parameter weights.

\subsection{Search Space and Continuous Relaxation of Architecture Representation}
\label{subsec:DARTS}

To search for the network architecture, we first define the search space.
As shown in Figure~\ref{fig:DARTS-ASR}, the search space is a directed acyclic graph consisting of $K$ nodes $\{n_0, n_1, ..., n_K\}$, where $n_0$ is the input feature $\mathbf{X}$ and the other nodes represent latent features $\mathbf{H_1}, \mathbf{H_2}, ..., \mathbf{H_K}$.
In the scenario of ASR, the input feature $\mathbf{X}$ is a segment of acoustic features such as Mel-filterbanks, and latent features $\mathbf{H_i}$ have the shape like CNN feature maps.
For each node $n_i$, there are $i$ directed input edges $\{e_{i,0}, e_{i,1}..., e_{i,i-1}\}$, where each edge $e_{i,j}$ transforms $\mathbf{H_j}$ with some operation $g_{i,j}$.
The feature $H_i$ of each node $n_i$ is the summation of the operations of all its previous nodes as below.
\begin{align}
\mathbf{H_i} &= \Sigma_{j<i}\ \ g_{i,j}(\mathbf{H_j}), \\
where\ \ g_{i,j}(\mathbf{H_j}) &= \Sigma_{f \in \mathbb{F}} \dfrac{\exp(\alpha^f_{i,j})}{\Sigma_{{f'} \in \mathbb{F}} \exp(\alpha^{f'}_{i,j})}f(\mathbf{H_j}).
\label{eq:o_ij}
\end{align}
The operation $g_{i,j}$ is the weighted sum of a set of transformations $\mathbb{F}$.  Each transformation acts as a typical network layer like 3x3Conv, MaxPool2d or skip connection. 
Some of the transformations have parameter weights to be learned (for example, 3x3Conv), while some of them do not (for example, MaxPool2d, skip connection).
The transformation weights in an operation are parameterized by a vector $\mathbf{\alpha_{i,j}}$ of dimension $|\mathbb{F}|$.
The final output of searched architecture is the concatenation of all the latent features:
\begin{align}
\mathbf{O} = Concat(\mathbf{H_1}, \mathbf{H_2}, ..., \mathbf{H_K}).
\label{eq:output}
\end{align}

These variables $\mathbf{\alpha_{i,j}}$ is jointly trained with parameter weights directly by gradient descent.
If the weights $\mathbf{\alpha_{i,j}}$ are sparse, equation (\ref{eq:o_ij}) can be regraded as the selection of transformations used to connect node $n_i$ and $n_j$, so $\mathbf{\alpha_{i,j}}$ can be considered as controlling the network architecture.
Therefore, architecture search can be performed through learning the continuous variables $\{\alpha_{i,j}\}$.
With continuous relaxation of architecture representation by variables $\{\alpha_{i,j}\}$, the transformation components and connections of the model can be softly designed by gradient descent optimization.

\subsection{Multi-lingual Pre-training and Adaptation}
\label{subsec:multi-lingual}
To examine the generalizability of DARTS-ASR, we apply DARTS-ASR on not only monolingual but also multilingual ASR to check if it works on ASR of different languages.
For monolingual ASR, each language data is separately trained with respective training data, and the model is not shared across languages.
For multilingual ASR, some source languages are used for pre-training and some target languages for adaptation.
For each source language in pre-training, the input is encoded by the shared model, and then fed into the language-specific head of the corresponding language to output the prediction sequence.
During adaptation of target languages, the pre-trained shared model is used for fine-tuning, but the head is trained from scratch.

We apply three types of fine-tuning approaches:
\begin{itemize}
\item Adapt only param.: the continuous variables $\{\alpha_{i,j}\}$ from pre-training are fixed, and only parameter weights in the transformations are trained.
That is, the network architecture is learned from the source languages, and with the learned architecture, its network parameters are learned from the target language. 
\item Adapt arch.+param.: the continuous variables $\{\alpha_{i,j}\}$  keep being trained with parameter weights in the transformations.
That is, both the network architecture and network parameters learned from source languages are further fined-tuned on the target language. 
\item Adapt pruned arch.+param.: the architecture learned from the source languages is pruned by removing some transformations with low $\{\alpha^f_{i,j}\}$ values.
Then the pruned $\{\alpha_{i,j}\}$ keeps being trained with remaining parameter weights.
\end{itemize}

\section{Experiments}
\label{sec:exp}

\subsection{Data and Features}
\label{subsec:data}

We conducted experiments on the Full Language Pack from the multilingual dataset, IARPA BABEL~\cite{gales2014speech}.
Three source languages were selected for multilingual pre-training: Bengali (Bn), Tagalog (Tl) and Zulu (Zu), and four target languages for adaptation: Vietnamese (Vi), Swahili (Sw), Tamil (Ta) and Kurmanji (Ku).
We followed the ESPnet recipe~\cite{watanabe2018espnet} for data preprocessing and final score evaluation.
The acoustic features are 80-dimensional Mel-filterbanks that are computed over a 25ms window every 10ms, plus 3-dimensional pitch features.

\subsection{Implementation Details}
\label{subsec:details}

Following the previous works~\cite{dalmia2018sequence,hsu2019meta}, we used a CNN-BiLSTM-Head structure as the multilingual ASR model, as shown in Figure~\ref{fig:model}(a), and adopted Connectionist Temporal Classification (CTC)~\cite{graves2006connectionist} loss as the objective function.
The baseline model architecture followed the previous work~\cite{hsu2019meta}, where the CNN module was a 6-layer VGG block as shown in Figure~\ref{fig:model}(b), and the BiLSTM module was a 3-layer bidirectional LSTM network with 360 cells in each direction.
We experimented with the channel number of convolutions in VGG as 128 or 512, and the results of these two settings in the following subsection were named as VGG-Small and VGG-Large.
The head used for each language was a linear matrix with softmax activation.

In this paper, we applied DARTS-ASR on the CNN module to search for a better architecture for extracting useful features from input.
To match the depth and the parameter size of VGG-Large, the number of nodes $K$ in the search space of DARTS-ASR, as mentioned in Subsection~\ref{subsec:DARTS}, was set to 5, and the channel number of convolutions were 32.
The transformation candidates in $\mathbb{F}$ were \{3x3 convolution, 5x5 convolution, 3x3 dilated convolution, 5x5 dilated convolution, 3x3 average pooling, 3x3 max pooling, skip connection\}.

In addition to standard convolution blocks and pooling, we also added dilated convolutions and skip connection into the transformation candidate set.
Dilated convolutions have generally improved the performance of semantic segmentation, as reported in a previous work~\cite{yu2015multi}.
The improvement comes from the fact that dilated convolutions expand the receptive field without loss of resolution or coverage.
Although convolutions with strides larger than one and pooling are similar concepts, both reduce the resolution.
Skip connection forwards the input to the next layer with an identity function and has been proved to avoid the problem of vanishing gradients. 
It has become very popular in recent CNN models such as DenseNet~\cite{huang2017densely} or ResNet~\cite{he2016deep}.
Therefore, these two types of transformations were also chosen as candidates during architecture search.

All transformations were of stride one (if applicable), and the convolved feature maps were padded to preserve their spatial resolution.
All convolutions were followed by ReLU activation and batch normalization~\cite{ioffe2015batch}.
The operation parametrization vectors $\mathbf{\alpha_{i,j}}$ described in Subsection~\ref{subsec:DARTS} were initialized as zero vectors to ensure equal amount of attention over all possible transformations, so parameter weights in every candidate transformation could receive sufficient gradients to learn at the beginning.
Adam~\cite{DBLP:journals/corr/KingmaB14} (lr=0.0001, betas=[0.5, 0.999], decay=0.001) was used as the optimizer for operation parametrization vectors $\mathbf{\alpha_{i,j}}$, and SGD (lr=0.01, momentum=0.9, decay=0.0003) was used as the optimizer for parameter weights.
The learning rate was reduced by a factor of 0.2 if no improvement for 3 epochs.
All of the training processes were terminated after the validation loss had converged.
The performances on the test sets were evaluated with greedy search decoding and 5-gram language model re-scoring.

\subsection{Results}
\label{results}

\subsubsection{Monolingual ASR}
\label{subsubsec:monoASR}

\begin{table}[thb]
\centering
\vspace{-0.3cm}
\caption{CER (\%) results of monolingual ASR using different CNN modules.}
\vspace{-0.3cm}
\label{table:mono}
\begin{tabular}{c||c|c|c|c}
\toprule[2pt]

\multicolumn{1}{c||}{\multirow{3}{*}{Language}} &
\multicolumn{4}{c}{CNN Module} \\ \cline{2-5}
 & VGG- & VGG- & \multicolumn{2}{c}{DARTS-ASR} \\ \cline{4-5}
 & Small & Large & Full & Only Conv3x3 \\ \toprule[2pt]
Vietnamese & 46.0 & 48.3 & \textbf{40.9} & 45.7 \\ \hline
Swahili & 39.6 & 38.3 & \textbf{35.9} & 36.8 \\ \hline
Tamil & 57.9 & 60.1 & \textbf{48.0} & 51.6 \\ \hline
Kurmanji & 57.2 & 56.8 & \textbf{55.5} & 56.5 \\
\toprule[2pt]
\end{tabular}
\vspace{-0.2cm}
\end{table}

\begin{figure}[t]
\vspace{-0.4cm}
  \centering
  \includegraphics[width=.9\linewidth]{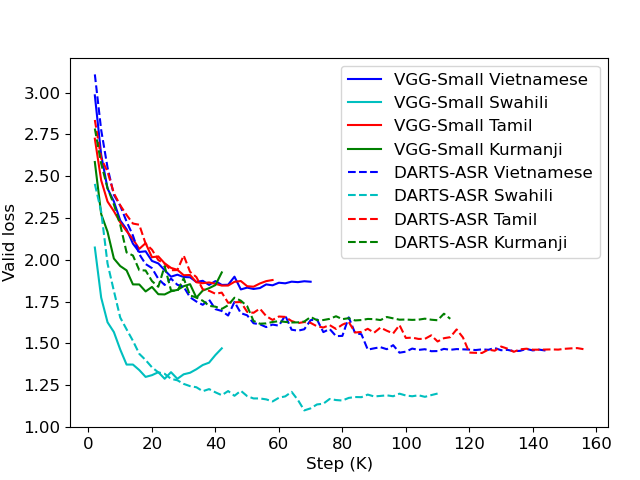}
  \vspace{-0.3cm}
  \caption{Validation loss vs training step with VGG-Small or DARTS-ASR for monolingual ASR on four languages.}
\label{fig:dev_loss}
\vspace{-0.5cm}
\end{figure}

For monolingual ASR on four languages, we evaluated diiferent kinds of CNN modules, VGG-Small and VGG-Large, as listed in Table~\ref{table:mono}.
The results of DARTS-ASR using all the seven kinds of transformations mentioned in the last subsection are listed in the third column.
We can observe DARTS-ASR significantly outperformed both VGG-Small and VGG-Large, showing that the connection pattern of nodes in DARTS-ASR contributed a lot to the huge performance boosting.
It is worth noting that even though the parameter size of VGG-Large was four times as many as VGG-Small, the CERs of Vietnamese and Tamil became worse due to overfitting and the CERs of Swahili and Kurmanji improved only a little.
In comparison, the parameter size of DARTS-ASR was also much larger than VGG-Small.
However, DARTS-ASR outperformed VGG-Small by 10.2\% relative reduction on average CER. It indicates the role of architecture for training DNN is very important.

To further understand the importance of the connection pattern and transformation candidates between nodes, in addition to the search space described in Subsection\ref{subsec:details}, we constructed another search space for DARTS-ASR: instead of having seven transformation candidates in the search space as described in Subsection~\ref{subsec:details}, there was only \{3x3 convolution\} in the search space.
The channel number of the convolution was set to 256 to match the parameter size of the original search space.
The results with only 3x3 convolution are listed in the fourth column.
DARTS-ASR outperformed VGG models even with limited search space.
It indicates the connection pattern of DARTS-ASR alone contributed a lot to performance improvement.
Furthermore, the performance of the full search space outperformed the \{3x3 convolution\} search space.
It proves that diversity of transformation candidates can provide the model an opportunity to find a better architecture.

In Figure~\ref{fig:dev_loss}, the validation losses of VGG-Small and DARTS-ASR on different languages are presented.
The solid lines are the results of VGG-Small and the dashed lines are those of DARTS-ASR.
Different colors stand for different languages.
From the lines, we can observe the convergence of VGG-Small was generally faster than DARTS-ASR.
But DARTS-ASR could reach much lower validation losses in the end.
The training of VGG-Small suffered from serious overfitting, causing the losses to increase again after some training steps.
In comparison, the validation losses of DARTS-ASR could decrease more steadily.

\begin{table}[thb]
\centering
\caption{CER (\%) results of multilingual ASR using DARTS-ASR under different fine-tuning approaches.}
\vspace{-0.3cm}
\label{table:fine-tune}
\begin{tabular}{c||c|c|c}
\toprule[2pt]

\multicolumn{1}{c||}{\multirow{3}{*}{Language}} &
\multicolumn{3}{c}{Fine-tuning of DARTS-ASR} \\ \cline{2-4}
 & Adapt & Adapt & Adapt pruned \\ 
 & only param. & arch.+param.  & arch.+param. \\ \toprule[2pt]
Vietnamese & \textbf{40.9} & \textbf{40.9} & 41.1 \\ \hline
Swahili & 33.2  & \textbf{32.3} & 35.3 \\ \hline
Tamil & 46.4  & \textbf{45.9} & 47.5 \\ \hline
Kurmanji & 53.6  & 53.5 & \textbf{53.2} \\
\toprule[2pt]
\end{tabular}
\vspace{-0.1cm}
\end{table}

\begin{table}[thb]
\centering
\caption{CER (\%) results of multilingual ASR using different CNN modules.}
\vspace{-0.3cm}
\label{table:multi}
\begin{tabular}{c||c|c|c}
\toprule[2pt]

\multicolumn{1}{c||}{\multirow{2}{*}{Language}} &
\multicolumn{3}{c}{CNN Module} \\ \cline{2-4}
 & \multirow{1}{*}{VGG-Small} & \multirow{1}{*}{VGG-Large} & DARTS-ASR \\ \toprule[2pt]
Vietnamese & 45.3 & 43.2 & \textbf{40.9} \\ \hline
Swahili & 36.3 & 36.1 & \textbf{32.3} \\ \hline
Tamil & 55.7 & 55.0 & \textbf{45.9} \\ \hline
Kurmanji & 54.5 & 55.1 & \textbf{53.5} \\ \toprule[2pt]
\end{tabular}
\vspace{-0.5cm}
\end{table}

\begin{figure*}[t]
\begin{minipage}[t]{0.16\linewidth}
  \centering
  \centerline{\includegraphics[width=\linewidth]{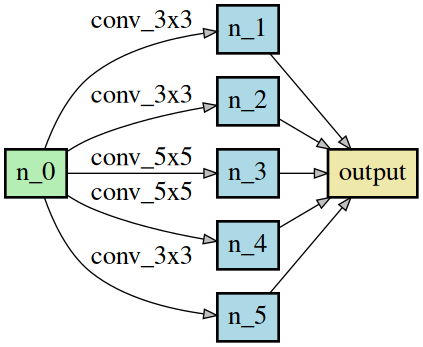}}
  \centerline{(a) Vietnamese.}\medskip
  \label{minipage:mono107} 
\end{minipage}
\hfill
\begin{minipage}[t]{0.16\linewidth}
  \centering
  \centerline{\includegraphics[width=\linewidth]{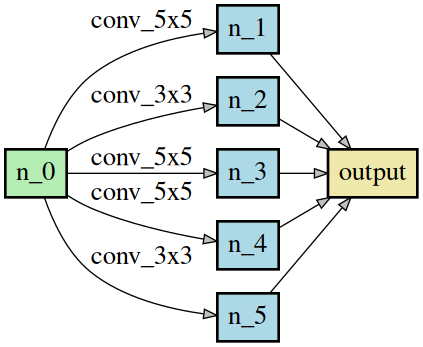}}
  \centerline{(b) Swahili.}\medskip
  \label{minipage:mono202} 
\end{minipage}
\hfill
\begin{minipage}[t]{0.30\linewidth}
  \centering
  \centerline{\includegraphics[width=\linewidth]{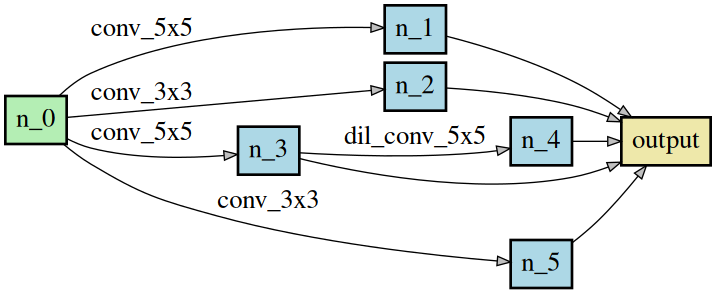}}
  \centerline{(c) Tamil.}\medskip
  \label{minipage:mono204} 
\end{minipage}
\hfill
\begin{minipage}[t]{0.36\linewidth}
  \centering
  \centerline{\includegraphics[width=\linewidth]{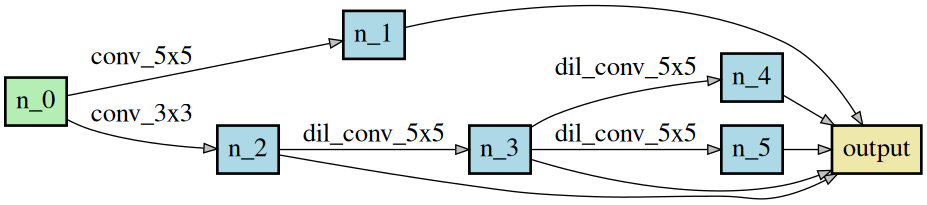}}
  \centerline{(d) Kurmanji.}\medskip
  \label{minipage:mono205} 
\end{minipage}
\vspace{-0.35cm}
\caption{Architectures for different languages found by DARTS-ASR in monolingual ASR.}
\label{fig:mono}
\vspace{-0.1cm}
\end{figure*}

\begin{figure*}[t]
\begin{minipage}[t]{0.49\linewidth}
  \centering
  \centerline{\includegraphics[width=\linewidth]{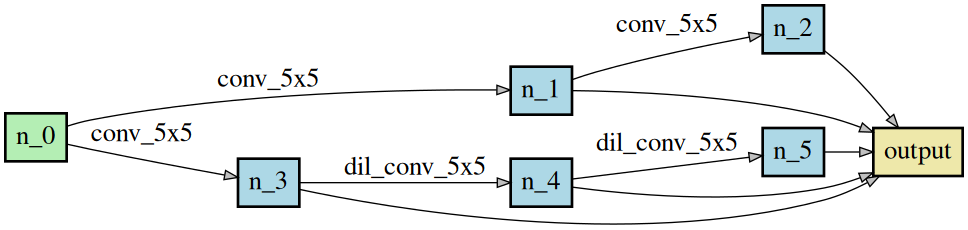}}
  \centerline{(a) Vietnamese and Kurmanji.}
  \label{minipage:multi107} 
\end{minipage}
\hfill
\begin{minipage}[t]{0.49\linewidth}
  \centering
  \centerline{\includegraphics[width=\linewidth]{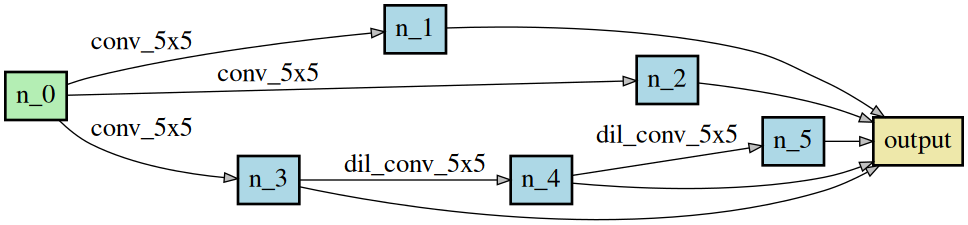}}
  \centerline{(b) Swahili and Tamil.}
  \label{minipage:multi202} 
\end{minipage}
\vspace{-0.2cm}
\caption{Architectures for different languages found by DARTS-ASR in multilingual ASR.}
\label{fig:multi}
\vspace{-0.3cm}
\end{figure*}

\subsubsection{Multilingual ASR}
\label{multiASR}

For multilingual ASR, the model was first pre-trained on three source languages mentioned in Subsection~\ref{subsec:data}, and then adapted on the same four different target languages as in the monolingual ASR experiments, respectively.

We first conducted experiments to compare the three fine-tuning approaches described in Subsection~\ref{subsec:multi-lingual}, as shown in Table~\ref{table:fine-tune}.
Especially for ``Adapt pruned arch.+params'', the architecture was pruned by removing all transformations but the top three ones with the highest $\{\alpha^f_{i,j}\}$ values in each edge. 
Then the pruned $\{\alpha_{i,j}\}$ kept being fine-tuned jointly with remaining parameter weights.

From Table~\ref{table:fine-tune}, we can observe ``Adapt arch.+param.'' fine-tuning approach obtained the best performance on average CER.
However, ``Adapt only param.'' and ``Adapt pruned arch.+param.'' were only a little worse than ``Adapt arch.+param.''.
It indicates after pre-training, DARTS-ASR can find a generally good architecture and parameter weights for different languages.
And the pruned architecture can reduce computational cost while suffering little performance drop.
We used ``Adapt arch.+param.'' fine-tuning approach for DARTS-ASR in the following experiments.

Then we compared DARTS-ASR with VGG-Small and VGG-Large.
The results are listed in Table~\ref{table:multi}.
All three kinds of CNN modules got much better performance on multilingual ASR than monolingual ASR.
On multilingual ASR, VGG-Large achieved better results than VGG-Small on average CER.
Among those, DARTS-ASR still outperformed both VGG-Small and VGG-Large by a significant margin.
It indicates DARTS-ASR can also benefit from multilingual learning to build a shared acoustic pre-trained model with a better architecture and parameter weights.

\subsubsection{Analysis of Searched Architectures}
\label{analysis}

We further plot and analyze the searched architectures by DARTS-ASR.
Similar to the original DARTS paper~\cite{liu2018darts}, to simplify the illustration of architecture, for each node $n_i$, we plot the most dominant transformation $f_j$ among all transformations in all entering edges. 
The selection of the most dominant transformation can be formulated as below.
\vspace{-0.15cm}
\begin{align}
f_j &= \arg\max_{j'<i}\ \alpha^{f_{j'}}_{i,j}, \\
where\ f_{j'} &= \arg\max_{f' \in \mathbb{F}}\ \alpha^{f'}_{i,j}.
\label{eq:max_input}
\end{align}
\vspace{-0.3cm}

The searched architecture for each language on monolingual ASR is shown in Figure~\ref{fig:mono}.
The architectures of Vietnamese and Swahili were similar, while those of Tamil and Kurmanji were quite different from one another.
For multilingual ASR, we plot the searched architectures under the ``Adapt arch.+params.'' fine-tuning approach.
The searched architectures of Vietnamese and Kurmanji were the same as shown in Figure~\ref{fig:multi}(a), and those of Swahili and Tamil were the same as shown in Figure~\ref{fig:multi}(b).
We can observe all of the four searched architectures on multilingual ASR were quite similar, where the patterns for nodes $n_3$ to $n_5$ in the bottom were all the same.
Only the patterns for $n_1$ and $n_2$ were slightly different.
It shows that this kind of network architecture shown in Figure~\ref{fig:multi} is the architecture generally suitable for a wide range of languages. 

\section{Conclusion}
\label{sec:conclusion}
In this paper, we propose an ASR approach with efficient gradient-based architecture search, DARTS-ASR.
In order to examine the generalizability of DARTS-ASR, we apply our approach not only on many languages to perform monolingual ASR, but also on a multilingual ASR setting.
The experiment results show that our approach outperformed the baseline fixed-topology architecture significantly under both monolingual and multilingual ASR settings.
Furthermore, we perform some analysis on the searched architectures by DARTS-ASR.
In future work, DARTS-ASR can be incorporated with other ASR or meta-learning approaches for further improvement.

\bibliographystyle{IEEEtran}

\begin{thebibliography}{10}
\providecommand{\url}[1]{#1}
\csname url@samestyle\endcsname
\providecommand{\newblock}{\relax}
\providecommand{\bibinfo}[2]{#2}
\providecommand{\BIBentrySTDinterwordspacing}{\spaceskip=0pt\relax}
\providecommand{\BIBentryALTinterwordstretchfactor}{4}
\providecommand{\BIBentryALTinterwordspacing}{\spaceskip=\fontdimen2\font plus
\BIBentryALTinterwordstretchfactor\fontdimen3\font minus
  \fontdimen4\font\relax}
\providecommand{\BIBforeignlanguage}[2]{{%
\expandafter\ifx\csname l@#1\endcsname\relax
\typeout{** WARNING: IEEEtran.bst: No hyphenation pattern has been}%
\typeout{** loaded for the language `#1'. Using the pattern for}%
\typeout{** the default language instead.}%
\else
\language=\csname l@#1\endcsname
\fi
#2}}
\providecommand{\BIBdecl}{\relax}
\BIBdecl

\bibitem{krizhevsky2012imagenet}
A.~Krizhevsky, I.~Sutskever, and G.~E. Hinton, ``Imagenet classification with
  deep convolutional neural networks,'' in \emph{Advances in neural information
  processing systems}, 2012, pp. 1097--1105.

\bibitem{he2016deep}
K.~He, X.~Zhang, S.~Ren, and J.~Sun, ``Deep residual learning for image
  recognition,'' in \emph{Proceedings of the IEEE conference on computer vision
  and pattern recognition}, 2016, pp. 770--778.

\bibitem{devlin2019bert}
J.~Devlin, M.-W. Chang, K.~Lee, and K.~Toutanova, ``Bert: Pre-training of deep
  bidirectional transformers for language understanding,'' in \emph{Proceedings
  of the 2019 Conference of the North American Chapter of the Association for
  Computational Linguistics: Human Language Technologies, Volume 1 (Long and
  Short Papers)}, 2019, pp. 4171--4186.

\bibitem{fukushima1980neocognitron}
K.~Fukushima, ``Neocognitron: A self-organizing neural network model for a
  mechanism of pattern recognition unaffected by shift in position,''
  \emph{Biological cybernetics}, vol.~36, no.~4, pp. 193--202, 1980.

\bibitem{DBLP:conf/iclr/ZophL17}
B.~Zoph and Q.~V. Le, ``Neural architecture search with reinforcement
  learning,'' in \emph{5th International Conference on Learning
  Representations, {ICLR} 2017}.

\bibitem{zoph2018learning}
B.~Zoph, V.~Vasudevan, J.~Shlens, and Q.~V. Le, ``Learning transferable
  architectures for scalable image recognition,'' in \emph{Proceedings of the
  IEEE conference on computer vision and pattern recognition}, 2018, pp.
  8697--8710.

\bibitem{real2019regularized}
E.~Real, A.~Aggarwal, Y.~Huang, and Q.~V. Le, ``Regularized evolution for image
  classifier architecture search,'' in \emph{Proceedings of the aaai conference
  on artificial intelligence}, vol.~33, 2019, pp. 4780--4789.

\bibitem{baruwa2019leveraging}
A.~Baruwa, M.~Abisiga, I.~Gbadegesin, and A.~Fakunle, ``Leveraging end-to-end
  speech recognition with neural architecture search,'' \emph{arXiv preprint
  arXiv:1912.05946}, 2019.

\bibitem{veniat2019stochastic}
T.~V{\'e}niat, O.~Schwander, and L.~Denoyer, ``Stochastic adaptive neural
  architecture search for keyword spotting,'' in \emph{ICASSP 2019-2019 IEEE
  International Conference on Acoustics, Speech and Signal Processing
  (ICASSP)}.\hskip 1em plus 0.5em minus 0.4em\relax IEEE, 2019, pp. 2842--2846.

\bibitem{mazzawi2019improving}
H.~Mazzawi, X.~Gonzalvo, A.~Kracun, P.~Sridhar, N.~Subrahmanya, I.~L. Moreno,
  H.~J. Park, and P.~Violette, ``Improving keyword spotting and language
  identification via neural architecture search at scale,'' \emph{Proc.
  Interspeech 2019}, pp. 1278--1282, 2019.

\bibitem{liu2018darts}
H.~Liu, K.~Simonyan, and Y.~Yang, ``{DARTS}: Differentiable architecture
  search,'' in \emph{International Conference on Learning Representations},
  2019.

\bibitem{wu2019fbnet}
B.~Wu, X.~Dai, P.~Zhang, Y.~Wang, F.~Sun, Y.~Wu, Y.~Tian, P.~Vajda, Y.~Jia, and
  K.~Keutzer, ``Fbnet: Hardware-aware efficient convnet design via
  differentiable neural architecture search,'' in \emph{Proceedings of the IEEE
  Conference on Computer Vision and Pattern Recognition}, 2019, pp.
  10\,734--10\,742.

\bibitem{xie2018snas}
\BIBentryALTinterwordspacing
S.~Xie, H.~Zheng, C.~Liu, and L.~Lin, ``{SNAS}: stochastic neural architecture
  search,'' in \emph{International Conference on Learning Representations},
  2019. [Online]. Available: \url{https://openreview.net/forum?id=rylqooRqK7}
\BIBentrySTDinterwordspacing

\bibitem{chen2019progressive}
X.~Chen, L.~Xie, J.~Wu, and Q.~Tian, ``Progressive differentiable architecture
  search: Bridging the depth gap between search and evaluation,'' in
  \emph{Proceedings of the IEEE International Conference on Computer Vision},
  2019, pp. 1294--1303.

\bibitem{vu2014multilingual}
N.~T. Vu, D.~Imseng, D.~Povey, P.~Motlicek, T.~Schultz, and H.~Bourlard,
  ``Multilingual deep neural network based acoustic modeling for rapid language
  adaptation,'' in \emph{2014 IEEE international conference on acoustics,
  speech and signal processing (ICASSP)}.\hskip 1em plus 0.5em minus
  0.4em\relax IEEE, 2014, pp. 7639--7643.

\bibitem{tong2017investigation}
S.~Tong, P.~N. Garner, and H.~Bourlard, ``An investigation of deep neural
  networks for multilingual speech recognition training and adaptation,'' in
  \emph{Proc. of INTERSPEECH}, no. CONF, 2017.

\bibitem{cho2018multilingual}
J.~Cho, M.~K. Baskar, R.~Li, M.~Wiesner, S.~H. Mallidi, N.~Yalta, M.~Karafiat,
  S.~Watanabe, and T.~Hori, ``Multilingual sequence-to-sequence speech
  recognition: architecture, transfer learning, and language modeling,'' in
  \emph{2018 IEEE Spoken Language Technology Workshop (SLT)}.\hskip 1em plus
  0.5em minus 0.4em\relax IEEE, 2018, pp. 521--527.

\bibitem{tong2017multilingual}
S.~Tong, P.~N. Garner, and H.~Bourlard, ``Multilingual training and
  cross-lingual adaptation on ctc-based acoustic model,'' \emph{arXiv preprint
  arXiv:1711.10025}, 2017.

\bibitem{yi2018adversarial}
J.~Yi, J.~Tao, Z.~Wen, and Y.~Bai, ``Adversarial multilingual training for
  low-resource speech recognition,'' in \emph{2018 IEEE International
  Conference on Acoustics, Speech and Signal Processing (ICASSP)}.\hskip 1em
  plus 0.5em minus 0.4em\relax IEEE, 2018, pp. 4899--4903.

\bibitem{adams2019massively}
O.~Adams, M.~Wiesner, S.~Watanabe, and D.~Yarowsky, ``Massively multilingual
  adversarial speech recognition,'' in \emph{Proceedings of the 2019 Conference
  of the North American Chapter of the Association for Computational
  Linguistics: Human Language Technologies, Volume 1 (Long and Short Papers)},
  2019, pp. 96--108.

\bibitem{sanabria2018hierarchical}
R.~Sanabria and F.~Metze, ``Hierarchical multitask learning with ctc,'' in
  \emph{2018 IEEE Spoken Language Technology Workshop (SLT)}.\hskip 1em plus
  0.5em minus 0.4em\relax IEEE, 2018, pp. 485--490.

\bibitem{dalmia2018sequence}
S.~Dalmia, R.~Sanabria, F.~Metze, and A.~W. Black, ``Sequence-based
  multi-lingual low resource speech recognition,'' in \emph{2018 IEEE
  International Conference on Acoustics, Speech and Signal Processing
  (ICASSP)}.\hskip 1em plus 0.5em minus 0.4em\relax IEEE, 2018, pp. 4909--4913.

\bibitem{hsu2019meta}
J.-Y. Hsu, Y.-J. Chen, and H.-y. Lee, ``Meta learning for end-to-end
  low-resource speech recognition,'' \emph{arXiv preprint arXiv:1910.12094},
  2019.

\bibitem{gales2014speech}
M.~J. Gales, K.~M. Knill, A.~Ragni, and S.~P. Rath, ``Speech recognition and
  keyword spotting for low-resource languages: Babel project research at
  cued,'' in \emph{Spoken Language Technologies for Under-Resourced Languages},
  2014.

\bibitem{watanabe2018espnet}
S.~Watanabe, T.~Hori, S.~Karita, T.~Hayashi, J.~Nishitoba, Y.~Unno, N.-E.~Y.
  Soplin, J.~Heymann, M.~Wiesner, N.~Chen \emph{et~al.}, ``Espnet: End-to-end
  speech processing toolkit,'' \emph{Proc. Interspeech 2018}, pp. 2207--2211,
  2018.

\bibitem{graves2006connectionist}
A.~Graves, S.~Fern{\'a}ndez, F.~Gomez, and J.~Schmidhuber, ``Connectionist
  temporal classification: labelling unsegmented sequence data with recurrent
  neural networks,'' in \emph{Proceedings of the 23rd international conference
  on Machine learning}, 2006, pp. 369--376.

\bibitem{yu2015multi}
F.~Yu and V.~Koltun, ``Multi-scale context aggregation by dilated
  convolutions,'' \emph{arXiv preprint arXiv:1511.07122}, 2015.

\bibitem{huang2017densely}
G.~Huang, Z.~Liu, L.~Van Der~Maaten, and K.~Q. Weinberger, ``Densely connected
  convolutional networks,'' in \emph{Proceedings of the IEEE conference on
  computer vision and pattern recognition}, 2017, pp. 4700--4708.

\bibitem{ioffe2015batch}
S.~Ioffe and C.~Szegedy, ``Batch normalization: Accelerating deep network
  training by reducing internal covariate shift,'' in \emph{International
  Conference on Machine Learning}, 2015, pp. 448--456.

\bibitem{DBLP:journals/corr/KingmaB14}
D.~P. Kingma and J.~Ba, ``Adam: {A} method for stochastic optimization,'' in
  \emph{3rd International Conference on Learning Representations, {ICLR} 2015},
  2015.

\end{thebibliography}

\end{document}